\documentclass[a4paper,11pt]{article}

\usepackage{amsmath,amssymb,mathtools}
\usepackage{color}
\usepackage{graphicx}
\usepackage{subfigure}
\usepackage{cite}
\usepackage[colorlinks=true,linkcolor=red,citecolor=blue,urlcolor=blue,bookmarks]{hyperref}
\usepackage{multirow,makecell}
\usepackage{textcomp}
\usepackage{wasysym}
\usepackage{ulem}

\usepackage{verbatim}

\usepackage[utf8]{inputenc}
\usepackage[T1]{fontenc}

\usepackage{array}

\usepackage{diagbox}  % diagonal box in a table

\usepackage[text={17.1cm,24.6cm},centering]{geometry} %% thanks to Chao Wu
\numberwithin{equation}{section}

\def \be {\begin{equation}}
\def \ee {\end{equation}}
\def \ba {\begin{array}}
\def \ea {\end{array}}
\def \bea {\begin{eqnarray}}
\def \eea {\end{eqnarray}}
\def \nn {\nonumber}

\def \a {\alpha}
\def \b {\beta}

\def \d {\delta}

\def \Lam {\Lambda}

\def \r {\rho}

\def \cH {\mathcal H}

\def \cX {\mathcal X}

\def \f {\frac}

\def \lt {\left}
\def \rt {\right}

\def \td {\tilde}

\def \lag {\langle}
\def \rag {\rangle}

\def \ep {\mathrm{e}}
\def \ii {\mathrm{i}}

\def \tr {\textrm{tr}}

\def \and {{~\textrm{and}~}}

\def \per {\mathop{\textrm{per}}}

\begin{document}

\title{
\textbf{Additivity of disjoint interval entanglement in quasiparticle excited states}
}

\author{
Zhouhao Guo %${}^1$,
%M.~A.~Rajabpour${}^2$,
 and
Jiaju Zhang%${}^1$
\footnote{Corresponding author: jiajuzhang@tju.edu.cn}
}
\date{}
\maketitle
\vspace{-10mm}
\begin{center}
{\it
%${}^1$
Center for Joint Quantum Studies and Department of Physics, School of Science, Tianjin University,\\
135 Yaguan Road, Tianjin 300350, China\\
%${}^2$Instituto de Fisica, Universidade Federal Fluminense,\\
%Av.~Gal.~Milton Tavares de Souza s/n, Gragoat\'a, 24210-346, Niter\'oi, RJ, Brazil
}
\vspace{10mm}
\end{center}

\begin{abstract}

  We investigate mixed-state entanglement measures, namely reflected entropy, mutual information and logarithmic negativity, for two disjoint intervals in one-dimensional systems excited by a finite number of quasiparticles. While whole system is in a pure state, the two disjoint intervals are in a generically  mixed state. To address the problem that natural subsystem bases are generically non-orthonormal in such excited states, we use a general and efficient algorithm that computes these measures directly from the density matrix expressed in an arbitrary non-orthonormal basis. Applying this method to classical, bosonic, and fermionic quasiparticle excitations on a circle, we discover a universal additivity property: in the limit of large momentum differences, the mixed-state entanglement of a multi-quasiparticle state decomposes exactly into the sum of independent contributions. This additivity unifies the entanglement behavior across classical and quantum statistics, with the classical result emerging naturally as a special case. Our findings establish a robust computational framework for mixed-state entanglement in excited many-body systems and reveal a generic decoupling mechanism that governs entanglement distribution beyond the ground state.

\end{abstract}

\baselineskip 18pt
\thispagestyle{empty}
\newpage

\tableofcontents

\section{Introduction}

Quantitative measures of entanglement between subsystems of quantum systems are important in quantum information theory \cite{Plenio:2007zz}. For two subsystems $A,B$ in a pure state $|\psi\rag$, the entanglement entropy is calculated as the von Neumann entropy of the reduced density matrix (RDM) $\r_A=\tr_B|\psi\rag\lag\psi|$ \cite{Bennett:1995ra,Popescu:1996hm}
\be
S_A = - \tr_A ( \r_A \log \r_A ).
\ee
The mixed-state case is more complicated. In a pure one-dimensional system, two disjoint intervals \(A\) and \(B\) typically form a mixed state \(\rho_{AB}\). Quantifying the quantum entanglement between the two disjoint intervals in this setting is an intriguing challenge that has inspired various measures. In this paper, we consider three mixed state entanglement measures, namely reflected entropy, mutual information, and logarithmic negativity, for two disjoint intervals in one-dimensional systems excited by a finite number of quasiparticles.

From $\r_{AB}$, one introduces auxiliary systems $A',B'$ and constructs the canonical purification for $AB$
\be
|\sqrt{\r_{AB}}\rag = ( \sqrt{\r_{AB}} \otimes I_{A'B'} ) \sum_{i,\a} |\psi^A_i\psi^B_\a\psi^{A'}_i\psi^{B'}_\a\rag,
\ee
where $\{\psi^A_i\}$ and $\{\psi^B_\a\}$ are canonical bases of the Hilbert spaces $\cH_A$ and $\cH_B$, respectively. The reflected entropy is then obtained from the RDM $\r_{AA'}=\tr_{BB'}|\sqrt{\r_{AB}}\rag\lag\sqrt{\r_{AB}}|$ as \cite{Dutta:2019gen}
\be
S_R(A:B) = - \tr_{AA'} ( \r_{AA'} \log \r_{AA'} ).
\ee
From the RDMs $\r_A=\tr_B \r_{AB}$ and $\r_B=\tr_A \r_{AB}$, the mutual information is defined as \cite{Nielsen:2010oan}
\be
I(A:B) = - \tr_A ( \r_A \log \r_A ) - \tr_B ( \r_B \log \r_B ) + \tr_{AB} ( \r_{AB} \log \r_{AB} ).
\ee
The logarithmic negativity is given by \cite{Vidal:2002zz,Plenio:2005cwa}
\be
E_N(A:B) = \log \tr_{AB} | \r_{AB}^{T_B} |,
\ee
where the partial transpose $T_B$ acts as
\be
\big( |\psi^A_i\psi^B_\a \rag \lag \psi^A_j\psi^B_\b | \big)^{T_B} = |\psi^A_i\psi^B_\b \rag \lag \psi^A_j\psi^B_\a |.
\ee
These measures satisfy the inequality  \cite{Dutta:2019gen}
\be \label{inequality}
S_R(A:B) \geq I(A:B).
\ee
For general discussion, we denote the three entanglement measures collectively by $\cX$
\be
\cX(A:B), ~~ \cX=S_R,I,E_N.
\ee

The entanglement entropy of a single interval in many-body states with a finite number of quasiparticles has been intensively studied \cite{Pizorn:2012aut,Berkovits:2013mii,Molter:2014qsb,%
Castro-Alvaredo:2018dja,Castro-Alvaredo:2018bij,Castro-Alvaredo:2019irt,Castro-Alvaredo:2019lmj,%
Zhang:2020vtc,Zhang:2020dtd,Zhang:2021bmy}, revealing various interesting features. When the large system size limit, large energy limit, and large momentum difference conditions are all satisfied, the entanglement entropy can be interpreted from the distribution probabilities of classical particles \cite{Castro-Alvaredo:2018dja,Castro-Alvaredo:2018bij,Castro-Alvaredo:2019irt,Castro-Alvaredo:2019lmj}. For finite momentum differences, the entanglement entropy generally depends on whether the excited quasiparticles are bosonic or fermionic; in any case, quasiparticles with large momentum differences contribute independently to the entanglement entropy \cite{Zhang:2020vtc,Zhang:2020dtd,Zhang:2021bmy}. The double interval entanglement entropy and logarithmic negativity of quasiparticle excited states in the classical limit have also been studied \cite{Castro-Alvaredo:2018dja,Castro-Alvaredo:2018bij,Castro-Alvaredo:2019irt,Castro-Alvaredo:2019lmj}, with results interpretable via classical particle distribution probabilities. In this paper, we investigate the double interval reflected entropy, mutual information, and logarithmic negativity of quasiparticle excited states in the purely quantum limit.

\begin{figure}[tp]
\centering
\includegraphics[width=0.3\textwidth]{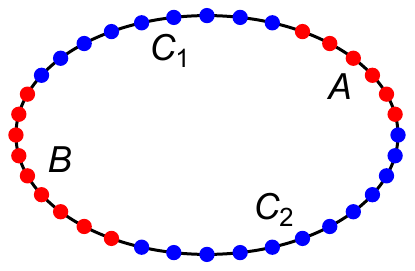}
\caption{The configuration of the subsystems $A=[1,\ell_1]$, $C_1=[\ell_1+1,\ell_1+d]$, $B=[\ell_1+d+1,\ell_1+d+\ell_2]$ and $C_2=[\ell_1+d+\ell_2+1,L]$. We define $C=C_1\cup C_2$.}
\label{FigureSubsystemAB}
\end{figure}

We consider a circular quantum system of $L$ sites, partitioned into subsystems $A=[1,\ell_1]$, $C_1=[\ell_1+1,\ell_1+d]$, $B=[\ell_1+d+1,\ell_1+d+\ell_2]$, and $C_2=[\ell_1+d+\ell_2+1,L]$, as shown in figure~\ref{FigureSubsystemAB}. Defining $C=C_1\cup C_2$, we study the dependence of entanglement measures on the fixed ratios
\be
x_1 = \f{\ell_1}{L}, ~~ x_2 = \f{\ell_2}{L}, ~~ y = \f{d}{L},
\ee
in the scaling limit $L,\ell_1,\ell_2,d \to +\infty$. We examine excited states of quasiparticles, which can be classical, bosonic or fermionic. Classical particles are understood as classical limits of bosonic or fermionic particles. A general excited state $|K\rag$ is labeled by the excited momenta
\be
|K\rag = \Big|\prod_{i=1}^s k_i^{r_i}\Big\rag,
\ee
where $k_i$ are the quasiparticle momenta. Using an algorithm in a non-orthonormal basis, we obtain numerical results for the double interval reflected entropy, mutual information, and logarithmic negativity. For a state $|K_1\cup K_2\rag$ with $K_1\cap K_2=\varnothing$ in the large momentum difference limit
\be \label{limit}
|k_1-k_2| \to \infty, ~ \forall k_1 \in K_1,\forall k_2 \in K_2,
\ee
we find an additivity property for all three measures
\be \label{additivity}
\cX_{K_1 \cup K_2} = \cX_{K_1} + \cX_{K_2}.
\ee
The known classical limit emerges as a special case of this additivity property. First, if only a single momentum of quasiparticle is excited, the results for bosons and fermions coincide with the classical result. Furthermore, when any two different excited quasiparticles have an infinite momentum difference, their contributions to the entanglement measure decouple, and the bosonic and fermionic results both approach the classical limit.

The remaining part of the paper is arranged as follows.
In section~\ref{sectionAlgorithm}, we detail the algorithm in a non-orthonormal basis.
In sections~\ref{sectionResults} we consider excited states of classical, bosonic and fermionic quasiparticles.
We conclude with discussions in section~\ref{sectionConclusion}.

\section{Calculation algorithm with non-orthonormal basis} \label{sectionAlgorithm}

We consider a tripartite system consisting of subsystems \(A\), \(B\) and \(C\), with Hilbert space dimensions \(d_A\), \(d_B\) and \(d_C\), respectively.
A general pure state of the whole system is expanded in a non-orthonormal product basis
\be
|\psi_{ABC}\rangle = \sum_{i,\alpha,a} \psi_{i,\alpha,a}\, | \phi^A_i \,\phi^B_\alpha \,\phi^C_a \rangle ,
\ee
where \(\{\phi^A_i,i=1,2,\cdots,d_A\}\), \(\{\phi^B_\alpha,\alpha=1,2,\cdots,d_B\}\) and \(\{\phi^C_a,a=1,2,\cdots,d_C\}\) are non-orthonormal bases for the three subsystems.
The reduced density matrix of the composite system \(AB\) is obtained by tracing out subsystem \(C\).
Let \([Q_C]_{a,b} = \langle \phi^C_a | \phi^C_b \rangle\) be the overlap matrix of the \(C\) basis.  The partial trace yields
\be
\rho_{AB} = \sum_{i,\alpha,j,\beta} P_{i,\alpha,j,\beta}\; | \phi^A_i \phi^B_\alpha \rangle\langle \phi^A_j \phi^B_\beta |,
\ee
with the coefficient tensor
\be
P_{i,\alpha,j,\beta} = \sum_{a,b} \psi_{i,\alpha,a}\, \psi^*_{j,\beta,b}\; [Q_C]_{a,b}.
\ee

In this section we present a robust and efficient algorithm that computes three fundamental mixed-state entanglement measures, reflected entropy \(S_R(A:B)\), mutual information \(I(A:B)\), and logarithmic negativity \(E_N(A:B)\), directly from the tensor \(\psi_{i,\alpha,a}\) and the overlap matrices of the non-orthonormal bases.  The algorithm works with the natural inner-product structure and avoids direct explicit orthogonalization, which reduces the computational complexity.  This framework is particularly well suited for studying entanglement in excited states of many-body systems, where non-orthonormal subsystem bases arise generically.

\subsection{Basis transformation to orthonormal basis}

We begin by defining the inner product matrices $Q_A$ and $Q_B$ with entries
\bea
&& [Q_A]_{i,j} = \lag  \phi^A_i | \phi^A_j \rag, \quad i,j=1,2,\cdots,d_A, \nn\\
&& [Q_B]_{\a,\b} = \lag  \phi^B_\a | \phi^B_\b \rag, \quad \a,\b=1,2,\cdots,d_B.
\eea
These positive definite Hermitian matrices can be diagonalized as
\be
Q_A = R_A \Lam_A R_A^\dag, \quad Q_B = R_B \Lam_B R_B^\dag,
\ee
where $R_A,R_B$ are unitary matrices and $\Lam_A,\Lam_B$ are diagonal matrices with positive entries. We define the composite matrices
\be
R = R_A \otimes R_B, \quad \Lam = \Lam_A \otimes \Lam_B.
\ee

The $d_A \times d_B \times d_A \times d_B$ tensor $P_{i,\a,j,\b}$ is reshaped into a $(d_A d_B) \times (d_A d_B)$ matrix
\be
P_{i\a,j\b} =  P_{i,\a,j,\b}.
\ee
We then construct the transformed matrix
\be
S = \sqrt{\Lam} R^\dag P R \sqrt{\Lam},
\ee
which yields the density matrix in orthonormal basis $\{|\psi^A_i\rag\}$, $\{|\psi^B_\a\rag\}$ as
\be \label{rAB}
\r_{AB} = \sum_{i,\a,j,\b} S_{i\a,j\b} | \psi^A_i \psi^B_\a \rag \lag \psi^A_j \psi^B_\b |.
\ee

\subsection{Reflected entropy calculation}

For the reflected entropy, we define the matrix square root
\be
T = \sqrt{S},
\ee
which is reshaped back to a $d_A \times d_B \times d_A \times d_B$ tensor
\be
T_{i,\a,j,\b} = T_{i\a,j\b}.
\ee
The canonical purification is then given by
\be
|\sqrt{\r_{AB}}\rag = \sum_{i,\a,j,\b} T_{i,\a,j,\b} | \psi^A_i \psi^B_\a \psi^{A'}_j \psi^{B'}_\b \rag.
\ee

We reorganize the tensor as $U_{i,j,\a,\b} = T_{i,\a,j,\b}$ and flatten it into a $d_A^2 \times d_B^2$ matrix
\be
U_{ij,\a\b} = U_{i,j,\a,\b}.
\ee
The reduced density matrix for $AA'$ becomes
\be
\r_{AA'} = \sum_{i,j,k,l} [UU^\dag]_{ij,kl} | \psi^A_i \psi^{A'}_j \rag \lag \psi^A_k \psi^{A'}_l |,
\ee
and the reflected entropy is calculated as
\be
S_R(A:B) = - \tr [ UU^\dag\log(UU^\dag) ].
\ee

\subsection{Mutual information calculation}

For mutual information, we reshape the matrix $S_{i\a,j\b}$ back to a tensor $S_{i,\a,j,\b} = S_{i\a,j\b}$. The reduced density matrices are obtained through partial traces
\be
[S_A]_{i,j} = \sum_\a S_{i,\a,j,\a}, \quad
[S_B]_{\a,\b} = \sum_i S_{i,\a,i,\b},
\ee
which give the RDMs in orthonormal basis as
\be \label{rArB}
\r_A = \sum_{i,j} [S_A]_{i,j} | \psi^A_i \rag \lag \psi^A_j |, \quad
\r_B = \sum_{\a,\b} [S_B]_{\a,\b} | \psi^B_\a \rag \lag \psi^B_\b |.
\ee
Using equations (\ref{rAB}) and (\ref{rArB}), the mutual information is
\be
I(A:B) = - \tr( S_A \log S_A ) - \tr( S_B \log S_B ) + \tr( S \log S ).
\ee

\subsection{Logarithmic negativity calculation}

For logarithmic negativity, we define the partially transposed tensor
\be
\td S_{i,\a,j,\b} = S_{i,\b,j,\a},
\ee
which is flattened into a $(d_A d_B) \times (d_A d_B)$ matrix
\be
\td S_{i\a,j\b} = \td S_{i,\a,j,\b}.
\ee
The partial transpose of the density matrix is then
\be
\r_{AB}^{T_B} = \sum_{i,\a,j,\b} \td S_{i\a,j\b} | \psi^A_i \psi^B_\a \rag \lag \psi^A_j \psi^B_\b |,
\ee
and the logarithmic negativity is given by
\be
E_N(A:B) = \log \tr | \td S |.
\ee

\subsection{Algorithm summary}

The complete computational procedure can be summarized as follows
\bea
&& \lt.\ba{r}
\lt.\ba{r}
\textrm{tensor } \psi \\
\textrm{matrix } Q_C
\ea\rt\}
\to \textrm{tensor } P \to \textrm{matrix } P \\
\textrm{matrices } Q_A,Q_B \to \textrm{matrices } R_A,R_B,\Lam_A,\Lam_B \to \textrm{matrices } R,\Lam
\ea\rt\}
\to \textrm{matrix } S \to \nn\\
&& \to \lt\{\ba{l}
\textrm{matrix } T \to \textrm{tensor } T \to \textrm{tensor } U \to \textrm{matrix } U \to S_R(A:B)  \\
\textrm{tensor } S \to
\lt\{\ba{l}
\textrm{matrices } S_A,S_B \to I(A:B) \\
\textrm{tensor } \td S \to \textrm{matrix } \td S \to E_N(A:B)
\ea\rt.
\ea\rt.
\eea

This algorithm provides a systematic approach for computing entanglement measures directly from the representation in non-orthonormal bases, which is particularly useful for numerical implementations in quantum many-body systems .

\section{Results of double-interval entanglement measures} \label{sectionResults}

This section presents the results for the double-interval entanglement measures in excited states of classical, bosonic and fermionic quasiparticles.

\subsection{Classical Particles} \label{sectionClassical}

We consider quantum states corresponding to distributions of classical particles, specifically quasiparticle excited states satisfying the large system limit, large energy limit, and large momentum difference limit \cite{Castro-Alvaredo:2018dja,Castro-Alvaredo:2018bij,Castro-Alvaredo:2019irt,Castro-Alvaredo:2019lmj}. For these states, the double-interval entanglement measures depend only on the ratios \(x_1\) and \(x_2\), and are independent of the separation \(y\).

\begin{figure}[tp]
\centering
\includegraphics[height=0.14\textheight]{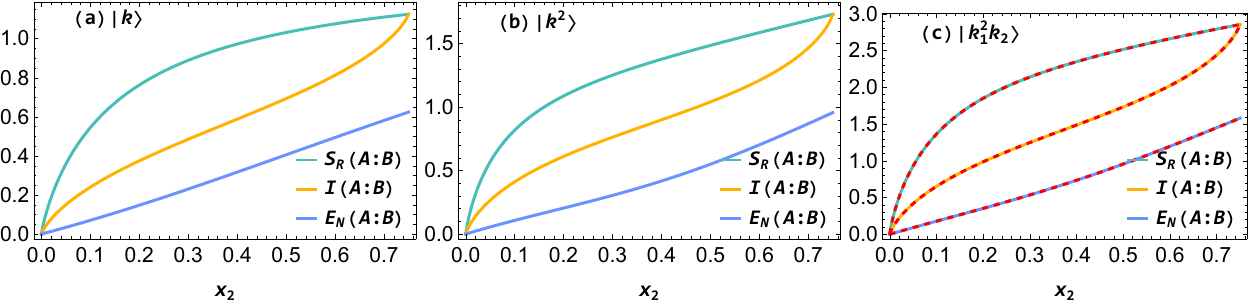}
\caption{
Dependence of reflected entropy, mutual information, and logarithmic negativity on \(x_2\) for states of identical (first and second panels) and different (third panel) classical particles. The red dashed line in the third panel shows the sum of contributions from different particles, \(\cX_{k^2}+\cX_{k}\) for \(\cX=S_R,I,E_N\). All panels have fixed \(x_1=\frac14\).
}
\label{FigureClassical}
\end{figure}

We begin with the one-particle state \(|k\rangle\), where \(k\) labels different classical particles and does not necessarily mean momentum here. Following the approach in \cite{Castro-Alvaredo:2019irt}, the density matrix is
\be
\rho_{AB} = (1-x_1-x_2) |00\rangle\langle 00| + x_2 |01\rangle\langle 01| + x_1 |10\rangle\langle 10| + \sqrt{x_1x_2}\,\bigl( |01\rangle\langle 10| + |10\rangle\langle 01| \bigr).
\ee
The symbol $|i_A i_B\rag$ denotes the orthonormal states with $i_A$ particles in $A$ and $i_B$ particles in $B$.
Applying the algorithm from section~\ref{sectionAlgorithm} yields the matrices
\be
S = \begin{pmatrix}
1-x_1-x_2 & 0              & 0              & 0 \\
0         & x_2            & \sqrt{x_1 x_2} & 0 \\
0         & \sqrt{x_1 x_2} & x_1            & 0 \\
0         & 0              & 0              & 0
\end{pmatrix},
\ee
\be
\tilde S = \begin{pmatrix}
1-x_1-x_2      & 0   & 0   & \sqrt{x_1 x_2} \\
0              & x_2 & 0   & 0 \\
0              & 0   & x_1 & 0 \\
\sqrt{x_1 x_2} & 0   & 0   & 0
\end{pmatrix},
\ee
\be
U = \begin{pmatrix}
\sqrt{1-x_1-x_2}        & 0                           & 0                           & \frac{x_2}{\sqrt{x_1+x_2}} \\[4pt]
0                       & 0                           & \sqrt{\frac{x_1 x_2}{x_1+x_2}} & 0 \\[4pt]
0                       & \sqrt{\frac{x_1 x_2}{x_1+x_2}} & 0                           & 0 \\[4pt]
\frac{x_1}{\sqrt{x_1+x_2}} & 0                           & 0                           & 0
\end{pmatrix}.
\ee
These lead to the analytical results
\bea \label{oneparticle}
&& S_R(A:B) = 2 h\!\Big(\frac{x_1x_2}{x_1+x_2}\Big)
+ \sum_{s=\pm1} h\!\Big( \frac{x_1+x_2-2x_1x_2+s\sqrt{(x_1+x_2)(x_1+x_2-4x_1x_2)}}{2(x_1+x_2)} \Big), \nn\\
&& I(A:B) = h(x_1) + h(1-x_1) + h(x_2) + h(1-x_2) - h(x_1+x_2) + h(1-x_1-x_2), \nn\\
&& E_N(A:B) = \log \bigl( x_1 + x_2 + \sqrt{(1-x_1-x_2)^2+4x_1x_2} \bigr),
\eea
with the function \(h(x)\) defined as
\be
h(x) = - x \log x .
\ee
The entanglement entropy and logarithmic negativity for a single-particle state were obtained in \cite{Castro-Alvaredo:2019irt}, while the reflected entropy result in this paper is new. We plot the results in panel (a) of figure~\ref{FigureClassical}.

For more general states \(|k^{r}\rangle\), examples of numerical results are shown in panel (b) of figure~\ref{FigureClassical}. The inequality \(S_R(A:B) \geq I(A:B)\) holds and saturates when \(AB\) is a pure state, yielding \(S_R(A:B)=I(A:B)=2S_A\). A positive Markov gap \(\Delta = S_R(A:B)-I(A:B)\) signals irreducible tripartite entanglement \cite{Hayden:2021gno}.

For states with distinct classical particles \(|k_1^{r_1}k_2^{r_2}\cdots\rangle\), panel (c) of figure~\ref{FigureClassical} shows that all three quantities \(\cX = S_R, I, E_N\) exhibit independent particle contributions
\be
\cX_{k_1^{r_1}k_2^{r_2}\cdots} = \cX_{k^{r_1}} + \cX_{k^{r_2}} + \cdots .
\ee
This additivity is expected for well-defined correlation measures, as distribution probabilities of distinct quasiparticles are independent.

\subsection{Bosonic Quasiparticles} \label{sectionBosonic}

We consider a free bosonic chain of $L$ sites, with each site $j=1,2,\cdots,L$ hosting a bosonic mode characterized by the lowering and raising operators $a_j$ and $a_j^\dag$, satisfying the commutation relation
\be
[a_{j_1},a_{j_2}^\dag]=\d_{j_1j_2}.
\ee
The ground state $|G\rag$ is defined as the state annihilated by all local lowering operators
\be
a_j |G\rag = 0, ~ j=1,2,\cdots,L.
\ee
For the subsystem configuration illustrated in figure~\ref{FigureSubsystemAB}, the ground state factorizes as a direct product $|G\rag=|G_A\rag\otimes|G_B\rag\otimes|G_C\rag$.

We introduce global momentum-space modes through the Fourier transformations
\be
b_k = \f{1}{\sqrt{L}} \sum_{j=1}^L \ep^{-\f{2\pi\ii j k}{L}} a_j, ~~
b_k^\dag = \f{1}{\sqrt{L}} \sum_{j=1}^L \ep^{\f{2\pi\ii j k}{L}} a_j^\dag,
\ee
with momentum index $k=1,2,\cdots,L$. The ground state is also annihilated by these global lowering operators:
\be
b_k |G\rag = 0, ~ k=1,2,\cdots,L.
\ee
For a set of momenta $K=\{k_1^{r_1},k_2^{r_2},\cdots,k_s^{r_s}\}$, a general globally excited state is constructed as
\be \label{generalstate}
|K\rag = \Big[ \prod_{i=1}^s \f{(b^\dag_{k_i})^{r_i}}{\sqrt{r_i!}} \Big] |G\rag.
\ee

Following \cite{Zhang:2020vtc,Zhang:2020dtd,Zhang:2021bmy}, we define subsystem modes for the partitions shown in figure~\ref{FigureSubsystemAB}
\be
b_{A,k}^\dag = \f{1}{\sqrt{L}} \sum_{j \in A} \ep^{-\f{2\pi\ii j k}{L}} a_j^\dag, \quad
b_{B,k}^\dag = \f{1}{\sqrt{L}} \sum_{j \in B} \ep^{-\f{2\pi\ii j k}{L}} a_j^\dag, \quad
b_{C,k}^\dag = \f{1}{\sqrt{L}} \sum_{j \in C} \ep^{-\f{2\pi\ii j k}{L}} a_j^\dag.
\ee
Note the relation $b_k^\dag=b_{A,k}^\dag+b_{B,k}^\dag+b_{C,k}^\dag$. For a momentum set $K=\{k_1^{r_1},k_2^{r_2},\cdots\}$, the corresponding subsystem raising operator is
\be
b_{A,K}^\dag = ( b_{A,k_1}^\dag )^{r_1} ( b_{A,k_2}^\dag )^{r_2} \cdots,
\ee
which generates the non-orthonormal basis $\{b_{A,K}^\dag |G_A\rag\}$ for subsystem $A$. The correlation matrix $Q_A$ has entries defined by
\be
[Q_A]_{K_1,K_2} = \begin{cases}
0                                                             & |K_1|\neq |K_2| \\
\mathop{\per}\limits_{k_1 \in K_1,k_2 \in K_2} \a_{A,k_1-k_2} & |K_1| = |K_2|
\end{cases},
\ee
where $|K_1|$ and $|K_2|$ denote the number of momenta in each set, $\per$ denotes the matrix permanent, and the coefficients $\a_{A,k}$ are given by
\be
\a_{A,k} = \begin{cases}
\f{\ell_1}{L} & k=0 \\
\ep^{-\f{\pi\ii k(\ell_1+1)}{L}} \f{\sin\f{\pi k\ell_1}{L}}{L\sin\f{\pi k}{L}} & k\neq0
\end{cases}.
\ee
Similarly, we define $b_{B,K}^\dag$ and obtain the non-orthonormal basis $\{b_{B,K}^\dag |G_B\rag\}$ for subsystem $B$, with correlation matrix $Q_B$ specified by
\be
\a_{B,k} = \begin{cases}
\f{\ell_2}{L} & k=0 \\
\ep^{-\f{\pi\ii k(2\ell_1+2d+\ell_2+1)}{L}} \f{\sin\f{\pi k\ell_2}{L}}{L\sin\f{\pi k}{L}} & k\neq0
\end{cases}.
\ee
For subsystem $C$, the basis $\{b_{C,K}^\dag |G_C\rag\}$ yields the correlation matrix $Q_C$ with
\be
\a_{C,k} = \d_{k,0} - \a_{A,k} - \a_{B,k}.
\ee

The general state (\ref{generalstate}) can be expanded in the non-orthonormal basis as
\be
b_{A,K_A}^\dag b_{B,K_B}^\dag b_{C,K_C}^\dag |G\rag.
\ee
The reduced density matrix $\r_{AB}$ is then expressed in the basis
\be
b_{A,K_A}^\dag b_{B,K_B}^\dag |G_{AB}\rag.
\ee
Using this formulation, we apply the algorithm from section~\ref{sectionAlgorithm} to compute the reflected entropy, mutual information, and logarithmic negativity.

\begin{figure}[tp]
\centering
\includegraphics[height=0.14\textheight]{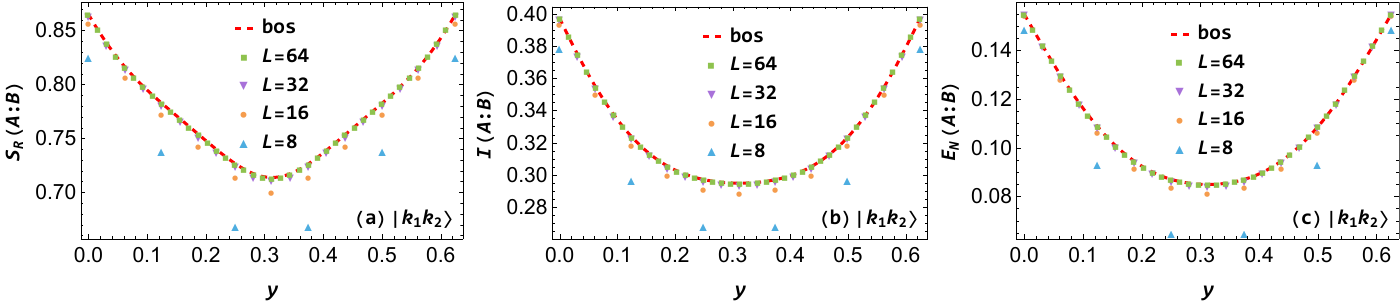}
\caption{Dependence of reflected entropy, mutual information, and logarithmic negativity on \(y\) for the state \(|k_1k_2\rangle\) with two different bosonic quasiparticles. Red dashed lines indicate results for \(L \to \infty\). All panels have fixed \((x_1,x_2)=(\frac18,\frac14)\) and \((k_1,k_2)=(1,2)\).}
\label{FigureBosonick1k2}
\end{figure}

For identical bosonic quasiparticle states
\be
|k^r\rangle = \frac{(b_k^\dag)^r}{\sqrt{r!}} |G\rangle,
\ee
the results match those for classical particles. The single-particle case yields identical results to (\ref{oneparticle}). For states with different quasiparticles such as \(|k_1k_2\rangle\), \(|k_1^2k_2\rangle\), and \(|k_1k_2k_3\rangle\), the results for fixed \(x_1,x_2\) depend on \(y\). In the scaling limit with fixed momenta, the results approach distinct forms termed bosonic results, as shown in figure~\ref{FigureBosonick1k2}.

\begin{figure}[tp]
\centering
\includegraphics[height=0.14\textheight]{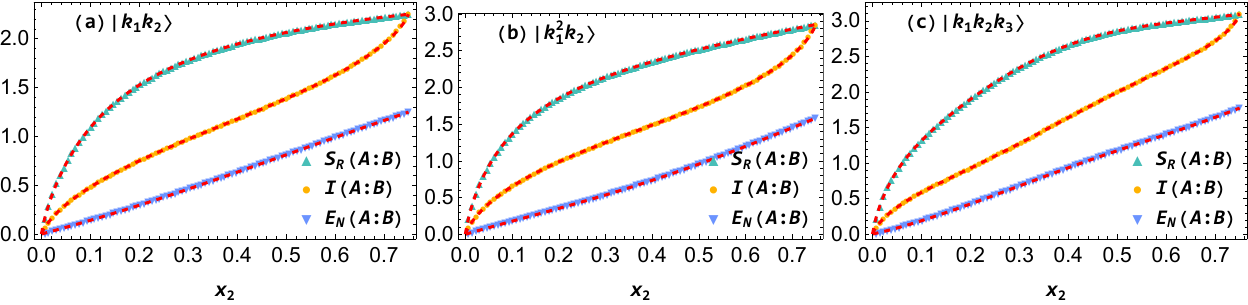}
\caption{Additivity of reflected entropy, mutual information, and logarithmic negativity in bosonic quasiparticle excited states. All panels have \(L=256\), \(x_1=\frac14\), and \(y=0\). The first and second panels have \((k_1,k_2)=(1,\frac{L}{4})\), the third panel has \((k_1,k_2,k_3)=(1,2,\frac{L}{4})\). Red dashed lines show predictions from additivity in the scaling limit \(L \to \infty\).}
\label{FigureBosonicAdditivity}
\end{figure}

For states \(|K_1 \cup K_2\rangle\) with \(K_1 \cap K_2 = \varnothing\) in the large momentum difference limit
\be
|k_1 - k_2| \to \infty, ~ \forall k_1 \in K_1, \forall k_2 \in K_2,
\ee
all three measures exhibit additivity
\be
\cX_{K_1 \cup K_2} = \cX_{K_1} + \cX_{K_2}.
\ee
Specifically, for the state
\be
|K\rangle = \Big|\prod_{i=1}^s k_i^{r_i}\Big\rangle
\ee
in the limit where all momentum differences are large, the result is a sum of classical contributions
\be
\cX_{k_1^{r_1}k_2^{r_2}\cdots} = \cX_{k^{r_1}} + \cX_{k^{r_2}} + \cdots.
\ee
Examples of this additivity are shown in figure~\ref{FigureBosonicAdditivity}.

\subsection{Fermionic Quasiparticles} \label{sectionFermionic}

We consider a free fermionic chain of length $L$, where each site $j = 1, 2, \cdots, L$ hosts a fermionic mode described by the creation and annihilation operators $a_j^\dag$ and $a_j$, respectively. These operators satisfy the canonical anticommutation relations
\be
\{ a_{j_1}, a_{j_2}^\dag \} = \d_{j_1 j_2}.
\ee
The global modes, defined in momentum space, are introduced via the Fourier transformations
\be
b_k = \frac{1}{\sqrt{L}} \sum_{j=1}^L e^{-\frac{2\pi i j k}{L}} a_j, \quad
b_k^\dag = \frac{1}{\sqrt{L}} \sum_{j=1}^L e^{\frac{2\pi i j k}{L}} a_j^\dag,
\ee
with momentum index $k = 1, 2, \cdots, L$.
For a set of distinct momenta $K = \{k_1, k_2, \cdots, k_r\}$, a general globally excited state is constructed as
\be
|K\rangle = \Big( \prod_{i=1}^r b^\dag_{k_i} \Big) |G\rangle,
\ee
where $|G\rangle$ denotes the ground state.

The calculations proceed similarly to those for fermionic quasiparticles in section~\ref{sectionBosonic}. A notable difference lies in the correlation matrix $Q_A$, and its entries are defined by
\be
[Q_A]_{K_1, K_2} = \begin{cases}
0 & |K_1| \neq |K_2| \\
\det\limits_{k_1 \in K_1, k_2 \in K_2} \, \alpha_{A, k_1 - k_2} & |K_1| = |K_2|
\end{cases}.
\ee
A similar formula applies to the correlation matrices $Q_B$ and $Q_C$.

\begin{figure}[tp]
\centering
\includegraphics[height=0.14\textheight]{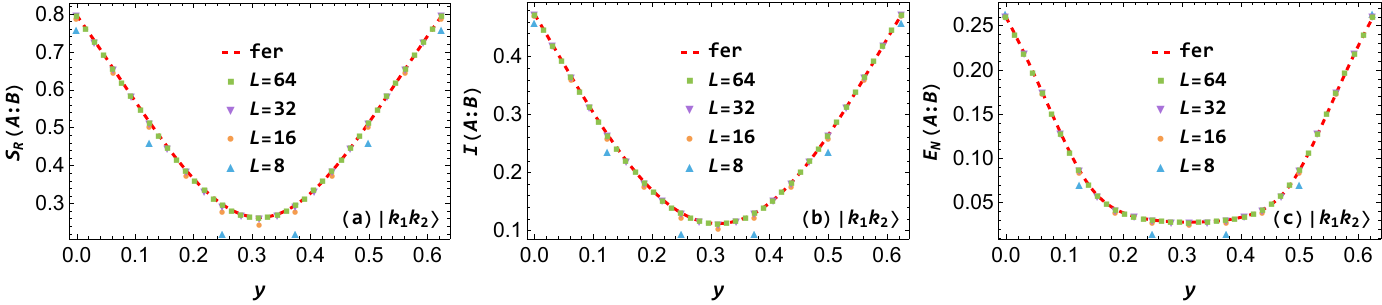}
\caption{Dependence of reflected entropy, mutual information, and logarithmic negativity on \(y\) for the state \(|k_1k_2\rangle\) with two different fermionic quasiparticles. Red dashed lines indicate results for \(L \to \infty\). All panels have fixed \((x_1,x_2)=(\frac18,\frac14)\) and \((k_1,k_2)=(1,2)\).}
\label{FigureFermionick1k2}
\end{figure}

For the single fermionic quasiparticle state
\be
|k\rangle = b_k^\dag |G\rangle,
\ee
the result matches the classical particle case (\ref{oneparticle}).
For states with different fermionic quasiparticles, the results for fixed \(x_1,x_2\) depend on \(y\). In the scaling limit with fixed momenta, the results approach distinct fermionic forms, as shown in figure~\ref{FigureFermionick1k2}.

\begin{figure}[tp]
\centering
\includegraphics[height=0.14\textheight]{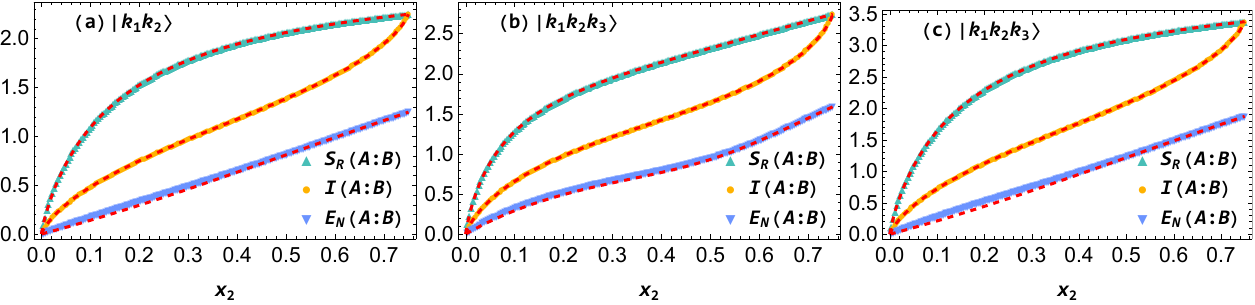}
\caption{Additivity of reflected entropy, mutual information, and logarithmic negativity in fermionic quasiparticle excited states. All panels have \(L=256\), \(x_1=\frac14\), and \(y=0\). The three panels correspond to \((k_1,k_2)=(1,\frac{L}{4})\), \((k_1,k_2,k_3)=(1,2,\frac{L}{4})\), and \((k_1,k_2,k_3)=(1,1+\frac{L}{4},1+\frac{L}{2})\), respectively. Red dashed lines show predictions from additivity in the scaling limit \(L \to \infty\).}
\label{FigureFermionicAdditivity}
\end{figure}

Similar to bosonic quasiparticles, states \(|K_1 \cup K_2\rangle\) with \(K_1 \cap K_2 = \varnothing\) in the large momentum difference limit exhibit the additivity property
\be
\cX_{K_1 \cup K_2} = \cX_{K_1} + \cX_{K_2}.
\ee
Specifically, for the state \(|k_1k_2\cdots k_r\rangle\) where all momentum differences are large, the result is
\be
\cX_{k_1k_2\cdots k_r} = r \cX_{k}.
\ee
Examples of this additivity are shown in figure~\ref{FigureFermionicAdditivity}.

\section{Conclusion and discussion} \label{sectionConclusion}

In this paper, we have developed an algorithm for calculating mixed-state entanglement measures, reflected entropy, mutual information, and logarithmic negativity, using a non-orthonormal basis. We applied this algorithm to investigate these double-interval entanglement measures in quasiparticle excited states, covering classical, bosonic, and fermionic cases. In the classical limit, the measures depend only on the subsystem sizes $x_1$ and $x_2$, and are independent of the separation $y$. For bosonic and fermionic quasiparticles, the entanglement measures exhibit a dependence on $y$ at finite momentum differences, but converge to a universal classical behavior in the large momentum difference limit. Similar to the single-interval entanglement entropy, we find that quasiparticles with large momentum differences contribute independently to the double-interval entanglement measures, a characteristic we refer to as the additivity property in the large momentum difference limit.

The reflected entropy was originally proposed in \cite{Dutta:2019gen} as a mixed-state entanglement measure and as a field theory dual of the entanglement wedge cross section in the bulk. In free field theory, its expected monotonic decrease under partial trace was verified in \cite{Bueno:2020vnx,Bueno:2020fle}. However, this monotonicity was later found to be violated in a specific classical state \cite{Hayden:2023yij} and in Lifshitz field theories \cite{Berthiere:2023bwn}, suggesting that reflected entropy may not always be a well-defined correlation measure. In this work, we have provided examples of states where monotonicity is satisfied. Establishing general criteria for the satisfaction or violation of monotonicity in broader classes of states remains an important direction for future research.

Our investigation has been confined to reflected entropy, mutual information, and logarithmic negativity in excited states with a finite number of quasiparticles within free bosonic and fermionic theories. It would be interesting to extend this study to other mixed-state entanglement measures, such as distillable entanglement and entanglement cost \cite{Plenio:2007zz}. Furthermore, examining the reflected entropy in more complex Gaussian states with an extensive number of quasiparticles in free bosonic and fermionic field theories, following approaches like \cite{Bueno:2020vnx,Bueno:2020fle}, could help determine whether the results from free theories apply to more general theories under appropriate limits, and whether the additivity property of entanglement measures holds broadly.
The additivity of single-interval entanglement entropy in the large-momentum-difference limit, originally observed in free theories \cite{Zhang:2020vtc,Zhang:2020dtd}, has also been shown to hold in interacting theories \cite{Zhang:2021bmy}. It would be interesting to investigate whether the additivity of double-interval entanglement measures found in free theories similarly extends to interacting models.

While the present work has concentrated on double-interval entanglement, the non-orthonormal basis algorithm we have developed applies directly to three or more intervals, thereby offering a practical computational tool for multipartite mixed-state entanglement measures. The characterization of genuine multipartite entanglement remains a fundamental open problem in quantum information theory \cite{Ma:2023ecg,Horodecki:2024bgc}. Quasiparticle excited states constitute an ideal testing ground for proposed measures: their entanglement structure is analytically tractable and can be systematically tuned by varying particle number, momenta, and subsystem geometry. In this controlled many-body setting, our method enables stable and efficient numerical evaluation of candidate measures, facilitating rigorous tests of their validity, monotonicity, and potential universal features.

\section*{Acknowledgements}

We thank M.~A.~Rajabpour for helpful discussions.
This work is supported by the National Natural Science Foundation of China (Grant No.~12205217) and the Tianjin University Self-Innovation Fund Extreme Basic Research Project (Grant No.~2025XJ21-0007).
Numerical calculations for this study were performed at high performance cluster at Center for Joint Quantum Studies (HPC-CJQS) of Tianjin University.

%\appendix

%\bibliographystyle{D:/00.bibx/JHEPx}
%\bibliography{D:/00.bibx/2025,D:/00.bibx/2024,D:/00.bibx/2023,D:/00.bibx/2022,D:/00.bibx/2021,D:/00.bibx/2020,D:/00.bibx/2019,D:/00.bibx/2018,D:/00.bibx/1960,D:/00.bibx/1970,D:/00.bibx/1980,D:/00.bibx/1990,D:/00.bibx/1995,D:/00.bibx/1996,D:/00.bibx/1997,D:/00.bibx/1998,D:/00.bibx/1999,D:/00.bibx/2000,D:/00.bibx/2001,D:/00.bibx/2002,D:/00.bibx/2003,D:/00.bibx/2004,D:/00.bibx/2005,D:/00.bibx/2006,D:/00.bibx/2007,D:/00.bibx/2008,D:/00.bibx/2009,D:/00.bibx/2010,D:/00.bibx/2011,D:/00.bibx/2012,D:/00.bibx/2013,D:/00.bibx/2014,D:/00.bibx/2015,D:/00.bibx/2016,D:/00.bibx/2017,D:/00.bibx/book,D:/00.bibx/work,D:/00.bibx/thesis}

\providecommand{\href}[2]{#2}\begingroup\raggedright\endgroup

\end{document}